\documentclass{jfm}

\usepackage{graphicx} 
\graphicspath{{./images/}}
\usepackage{amsmath,bm} 

\usepackage{xcolor}
\usepackage{soul} 
\usepackage{caption}
\usepackage{subcaption}
\usepackage{makecell} 

\usepackage{lineno}

\begin{document}

\shorttitle{Water waves generated by moving atmospheric pressure} 
\shortauthor{Liu and Higuera} 

\title{Water waves generated by moving atmospheric pressure: Theoretical analyses with applications to the 2022 Tonga event}

\author{Philip L.-F. Liu\aff{1}$^,$\aff{2}$^,$\aff{3}$^,$\aff{4} and Pablo Higuera\aff{1}$^,$\aff{5}}

\affiliation{
\aff{1}Department of Civil and Environmental Engineering, National University of Singapore, Singapore 
\aff{2}School of Civil and Environmental Engineering, Cornell University, Ithaca, 
USA
\aff{3}Institute of Hydrological and Oceanic Sciences, National Central University, 
Taiwan, 
\aff{4}Department of Hydraulic and Ocean Engineering, National Cheng Kung University, 
Taiwan, 
\aff{5}Department of Civil and Environmental Engineering, The University of Auckland, NZ
}

\maketitle

\begin{abstract}
Both 1DH (dispersive and non-dispersive) and 2DH axisymmetric (approximate, non-dispersive) analytical solutions are derived for water waves generated by moving atmospheric pressures.
In 1DH, three wave components can be identified: the locked wave propagating with the speed of the atmospheric pressure, $C_p$, and two free wave components propagating in opposite directions with the respective wave celerity, according to the linear frequency dispersion relationship. Under the supercritical condition ($C_p > C$, which is the fastest celerity of the water wave) the leading water wave is the locked wave and has the same sign (i.e., phase) as the atmospheric pressure, while 
the trailing free wave has the opposite sign.
Under the subcritical condition ($C >C_p$) the fastest moving free wave component leads and its free surface elevation has the same sign as the atmospheric pressure. For a long atmospheric pressure disturbance, the induced free surface profile mimics that of the atmospheric pressure. 
The 2DH problem involves an axisymmetric atmospheric pressure decaying in the radial direction as $O(r^{-1/2})$. Only two wave components, locked and free, appear due to symmetry.

The tsunami DART data captured during Tonga's volcanic eruption event is analyzed. Corrections are necessary to isolate the free surface elevation data.
Comparisons between the corrected DART data and the analytical solutions, including the arrival times of the leading locked waves and the trailing free waves, and the amplitude ratios, are in agreement in order-of-magnitude. The differences between them highlight the complexity of problems. 
\end{abstract}

\section{Introduction}

Atmospheric pressure variations or disturbances can be generated by a number of processes. The most common cause is a weather system (i.e., low pressure fronts, storms or hurricanes), which, in turn, produces a sea level anomaly, also known as storm surge \citep[e.g.,][]{bode97, pelinovsky01}.
Typically, low atmospheric pressure fronts propagate at a speed slower than the long water wave celerity. However, when these two speeds are close, especially over a shallow bathymetry, the Proudman resonance \citep{proudman29} may occur, producing larger surge responses, which are often called meteotsunamis \citep{monserrat06}.

Another source of atmospheric pressure disturbances is related to volcanic explosions. For example, the 1983 Krakatoa volcanic explosion in Indonesia\citep{harkrider67, garrett70} and the 2022 Hunga Tonga-Hunga Ha'apai underwater volcano explosion \citep[see][]{amores22, carvajal22} produced atmospheric pressure disturbances, which were captured by barometers all over the world. They also generated tsunami-like ocean waves. During the 2022 Tonga event, tsunami waves were reported across the Pacific Ocean and beyond, measured by the Deep-ocean Assessment and Reporting of Tsunamis (DART) system in deep water, and by tidal gauges placed at shallower coasts \citep{kataoka22}, including places not directly connected to the water body surrounding the Tonga volcano (e.g., Atlantic Ocean and Caribbean and Mediterranean Seas).

In both Krakatoa and Tonga events, the leading tsunami waves arrived much earlier than estimated by using the typical tsunami wave celerity ($\approx 200~$m/s). They were, instead, highly correlated with propagation speed of the atmospheric pressure waves \cite [][estimated 307 m/s]{amores22}. During the Tonga event, a second train of tsunami waves was later recorded by the sensors in the Pacific Ocean, propagating at the typical tsunami wave celerity. The trailing tsunami wave train has been attributed to other tsunami generation mechanisms associated with the volcano explosion and collapse \citep{lynett22}.

A clear distinction between the weather system generated storm surge/meteotsunami and the volcano explosion generated tsunamis is the relative speeds of the atmospheric pressure wave and the tsunami celerity. Typically, in the case of storm surges/meteotsunamis the former is slower than the latter, which can be called subcritical condition. For the volcanic explosion generated tsunamis, the opposite is true and is called the supercritical condition. The resulting tsunami wave characteristics are quite different. The objective of the paper is to use analytical solutions to better understand and further illustrate the relationships between the driving atmospheric pressure and the resulting water waves.

In this paper analytical solutions for water waves generated by moving atmospheric pressure are sought for 1DH (section~\ref{sec:dispSol}) and axisymmetric 2DH (section~\ref{sec:LSWAxisymmetricsol}) problems.
Small amplitude wave theory is adopted, while both dispersive and non-dispersive systems are considered.
To facilitate the analysis, constant depth is assumed.
In the 1DH problem the strength of the atmospheric pressure remains constant, but in the axisymmetric problem the pressure strength decays as $1/\sqrt{r}$, where $r$ measures the distance from the origin. 
The atmospheric pressure disturbances travel at different speeds relative to the long wave celerity and wave patterns/characteristics are investigated for both supercritical and subcritical conditions.
The analytical solutions are applied to the Tonga event data in section~\ref{sec:applicationPacific}.
Because of the presence of atmospheric pressure disturbances, a method to correct the reported DART system measurements for the free surface elevation is suggested.
The general characteristics of the tsunami waves are found to be captured by the analytical solutions.  
Finally, concluding remarks are provided in section {\ref{conclusion}}.

\section{1DH formulation and solutions}
\label{sec:dispSol}
Consider ocean waves generated by a prescribed atmospheric pressure field, $P_a(x,t)$, on the free surface, $z = \eta (x,t)$, in the two-dimensional vertical plane, $(x,z)$. Neglecting the viscous effects, the velocity potential, $\Phi (x,z,t)$, satisfies the continuity equation:
\begin{equation}
\frac{\partial^2\Phi}{\partial x^2} + \frac{\partial^2\Phi}{\partial z^2}=0. 
\end{equation}
The ocean bottom is approximated as a horizontal solid surface, $z=-h$. Thus, the no-flux boundary condition requires
\begin{equation}
    \frac{\partial \Phi}{\partial z} = 0 \text{ at } z=-h.
    \label{BBC}
\end{equation} 
Anticipating that the generated wave amplitude is small, the linearized free surface boundary conditions are applied on the still water surface ($z=0$) as
\begin{equation}
\frac{\partial \Phi}{\partial z} =\frac{\partial \eta}{\partial t} \;\; \text{and} \;\; \frac{\partial \Phi}{\partial t} + g\eta =-\frac{P_a(x,t)}{\rho} \text{ at } z=0, 
\end{equation}
 where $\rho$ is the density of water and $g$ is the gravity acceleration.  These two free surface boundary conditions can be combined by eliminating $\eta$, yielding 
\begin{equation}
\frac{\partial \Phi}{\partial z} =-\frac{1}{\rho g}\left [\rho \frac{\partial^2 \Phi}{\partial t^2} + \frac{\partial P_a}{\partial t}\right ] \text{ at } z=0. 
\label{CFSBC}
\end{equation}

In this paper $P_a$ is prescribed as a moving pressure field with a constant speed, $C_p$, starting at $t=0$. Thus, $P_a(x,t) = P_a(x - C_p t)$. 
Moreover, the wave motions begin from the quiescent state, i.e., $\eta (x, t=0^-) = \Phi (x,z,t= 0^-) =0$.

Applying the Laplace and Fourier transforms, namely
\begin{equation*}
\overline{\Phi} (x,z,s) =\int_0^\infty e^{-st}\Phi (x,z,t) dt, \;\; \text{and} \;\; \hat{\Phi}(k,z,t) =\frac{1}{\sqrt{2\pi}}\int_{-\infty}^{\infty} e^{-ikx}\Phi (x,z,t) dx,
\end{equation*}
to the initial boundary value problem stated above, the solutions for the transformed velocity potential and free surface elevation can be readily obtained as
\begin{equation}
    \hat{\overline{\Phi}}(k,z,s)=-\frac{\hat{P}_a (k)}{\rho}\left(\frac{s}{\omega^2+s^2}\right)\left(\frac{1}{s+ikC_p}\right)\frac{\cosh k(z+h)}{\cosh kh},
\label{potential_tran}
\end{equation}
\begin{equation}
\hat{\overline{\eta}}(k,s)=-\frac{\hat{P}_a (k)}{\rho g}\left(\frac{\omega^2}{\omega^2+s^2}\right)\left(\frac{1}{s+ikC_p}\right),
\label{free surface_tran}
\end{equation}
where $\omega^2 =gk \tanh kh$ is the dispersion relation and $\hat{P}_a (k)$ denotes the Fourier transform of $P_a$ at $t=0$. Applying the inverse Fourier and Laplace transforms to (\ref{potential_tran}) and (\ref{free surface_tran}), the velocity potential and free surface elevation can be obtained. Here, only the free surface elevation solution will be presented. 
The inverse Laplace transform on (\ref{free surface_tran}) will be performed first. There are three simple poles in (\ref{free surface_tran}), $s =\pm i \omega,$ and $-ikC_p$. Applying the Cauchy residual theorem, the inverse Laplace transform yields
\begin{equation*}
\hat{\eta} = -\frac{\hat{P}_a (k)}{\rho g}\left \{ 
    \left ( \frac{\omega^2}{\omega^2 -k^2C_p^2}\right ) e^{-ikC_pt}
    - \frac{1}{2}\left (\frac{\omega}{\omega -kC_p}\right ) e^{-i\omega t} 
    -\frac{1}{2}\left (\frac{\omega}{\omega +kC_p} \right ) e^{i\omega t}
    \right \}.
\end{equation*}
Now, applying the inverse Fourier transform to the equation above yields
\begin {linenomath}
\begin {align}
\begin {split}
    \eta (x,t) &=\eta_p +\eta_+ + \eta_-; \\
    \eta_p &=\frac{1}{\rho g} \frac{1}{\sqrt{2 \pi}}\int_{-\infty}^{\infty}\left ( \frac{C^2}{C_p^2 -C^2} \right )\hat{P}_a (k) e^{ik(x-C_pt)} dk ; \\
    \eta_+ &=\frac{1}{\rho g} \frac{1}{\sqrt{2 \pi}}\int_{-\infty}^{\infty}\frac{1}{2}\left ( \frac{C}{C -C_p} \right )\hat{P}_a (k) e^{ik(x-C t)} dk; \\
    \eta_- &=\frac{1}{\rho g} \frac{1}{\sqrt{2 \pi}}\int_{-\infty}^{\infty}\frac{1}{2}\left ( \frac{C}{C +C_p} \right )\hat{P}_a (k) e^{ik(x+C t)} dk,
    \label{solutions}
\end{split}
\end{align}
\end{linenomath}
where 
$C(k) =\omega/k$,
 represents the celerity of the generated wave component with wave number, $k$, satisfying the dispersion relation, $\omega^2=gk \tanh kh$. For a given water depth, $h$, the maximum celerity is $\sqrt{gh}$ as $k \rightarrow 0$.
 \begin{figure}
    \centering
    \includegraphics[width=0.70\linewidth]{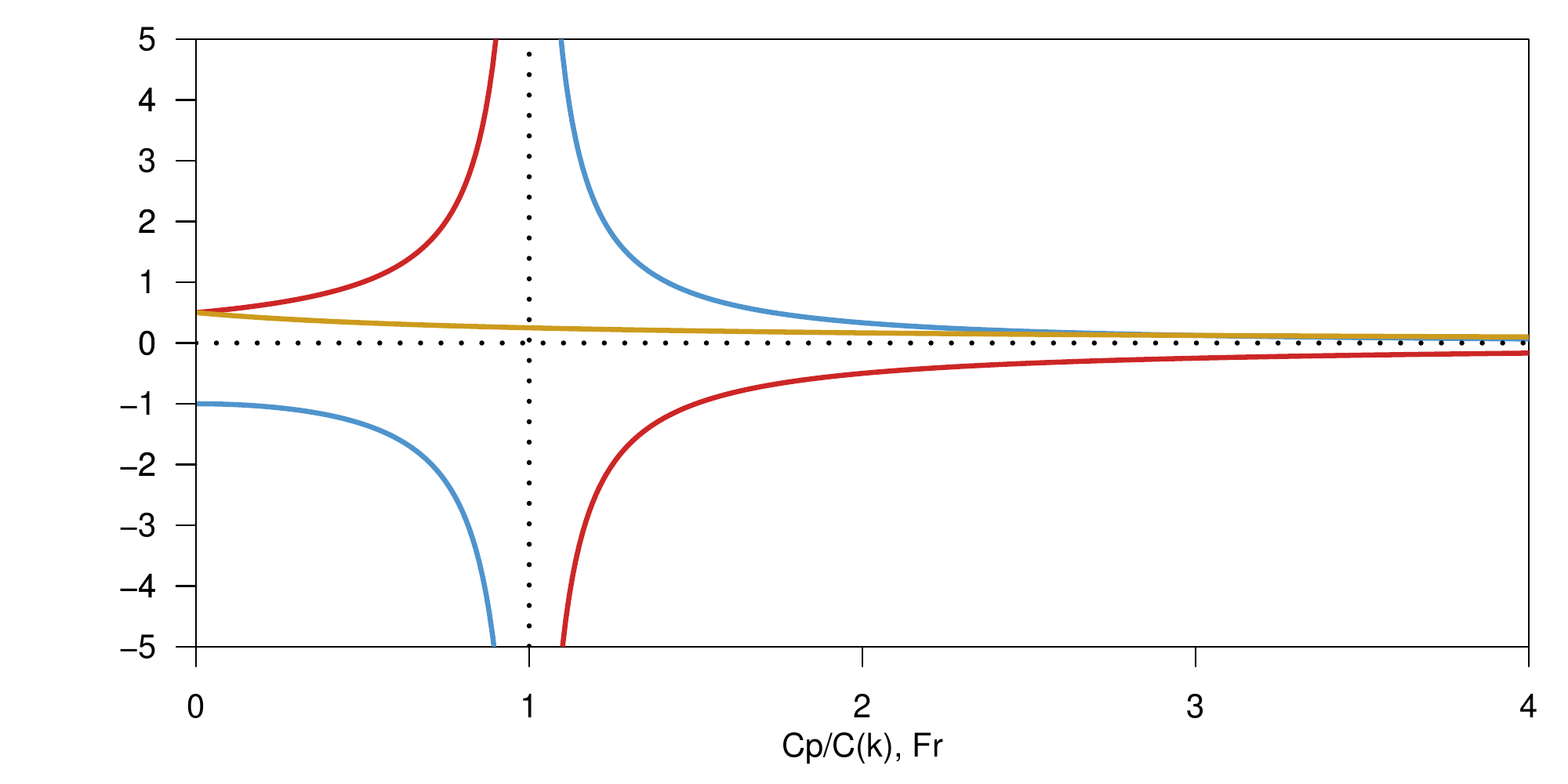}
    \caption{Modification functions of $\eta_p$ (blue line), $\eta_{+}$ (red line) and $\eta_{-}$ (orange line). The horizontal axis corresponds to $C_p/C(k)$ for the 1DH dispersive solution and to $F_r = C_p/\sqrt{g h}$ for the 1DH shallow water solution.}
    \label{fig:CCpAnalysis}
\end{figure}

The solutions given in (\ref{solutions}) are written in  integral forms, which can be numerically integrated once the atmospheric pressure and its Fourier transform, $\hat{P}_a$, are  provided. The first integral, $\eta_p$, represents a wave train, being ``locked'' with the moving atmospheric pressure with the propagation speed of $C_p$. The second and third integrals in (\ref{solutions}), $\eta_+$ and $\eta_-$, represent ``free'' waves, propagating in the $\pm$ $ x$-direction with the speed of $C(k)$, respectively. The shapes of these wave components are determined by the product of the atmospheric pressure spectral density, $\hat{P}_a(k)$, and a modification function. For the locked wave, the modification function is $C^2/(C_p^2 -C^2)$, while for the free waves, $\eta_+$ and $\eta_-$, the modification functions are, $C/(2(C -C_p))$ and $C/(2(C+C_p))$, respectively. 

These modification functions are plotted against $C_p/C(k)$ in figure~\ref{fig:CCpAnalysis}. Note that since $0 < C < \sqrt{gh}$, the applicable range of these curves for a given $C_p$ is $F_r < C_p/C < \infty$, where $F_r = C_p/\sqrt{gh}$ can be viewed as the Froude number of the problem. When the atmospheric wave (and the locked wave) propagates faster than the fastest free wave speed, $\sqrt{gh}$, it is called the supercritical condition ($F_r > 1$). The locked wave, $\eta_p$, is the leading wave moving in the $+ x$-direction.  On the other hand, $F_r <1$ is called the subcritical condition and the longest free wave component is the leading wave. For $F_r =1$, the propagation speed of the locked wave is the same as that of the fastest free wave, creating a resonance situation, which is called  the critical condition.

The sign and shape of free surface elevation depend on $\hat{P}_a$ over a range of $k$. From figure~\ref{fig:CCpAnalysis}, the modification function for $\eta_p$ is positive for $F_r >1$. Therefore, the locked wave free surface elevation has the same sign as that of the atmospheric pressure wave, although their shapes are not necessarily the same.
The modification function changes sign at $C_p/C =1$, which is an integrable singularity, and becomes negative for $F_r<1$.  
The modification function for the free wave, $\eta_+$, has the opposite sign of that for the locked wave, resulting in the opposite sign in free surface elevations. On the other hand, the modification function for the free wave, $\eta_-$, is always positive so that the free surface elevation has the same sign as that of the atmospheric pressure. Finally, the magnitude of the modification function for $\eta_-$ is relatively small as compared with those for the other two wave propagation modes, implying that the amplitude of the left-going free wave is also relatively small.

\subsection{Further analysis of the far field solution as $x \rightarrow \infty$}
\label{sec:analysisDispersive}

For a large time $t$, the most important contribution to the generated water waves comes from the long wave component, $k \approx 0$.
The locked wave in (\ref{solutions}) can be approximated as 
\begin {equation}
\eta_{p}  \approx  \frac{1}{\rho g}\left (\frac{gh}{C_p^2 - gh} \right )\frac{1}{\sqrt{2 \pi}}\int_{-\infty}^{\infty}\hat{P}_a (k) e^{ik(x-C_pt)} dk =\left (\frac{1}{F_r^2 - 1} \right )\frac{P_a (x-C_pt)}{\rho g}.
\label{solution_leading}
\end{equation}
Therefore, the free surface profile of the locked wave has the same shape as $P_a$. However, its magnitude is  multiplied by the modification factor, $1/(F_r^2- 1)$, which is also shown in figure~\ref{fig:CCpAnalysis}, with the horizontal axis being replaced by $F_r$.
In the supercritical regime ($F_r > 1$) the locked wave is the leading wave and this factor is positive (see the blue line in figure~ \ref{fig:CCpAnalysis}).
Therefore, the free surface profile and atmospheric pressure have the same sign, i.e., the positive atmospheric pressure induces the elevated (positive) free surface profile.
The modification factor becomes greater than one for $ F_r < \sqrt{2}$ and the amplitude of the locked wave diminishes to zero as $F_r \rightarrow \infty$ (i.e., the atmospheric pressure moves too fast for the water to respond).
In the critical condition ($F_r = 1$), resonance occurs as $ F_r \rightarrow 1$.
Under the subcritical condition ($F_r < 1$), the modification factor for $\eta_{p}$ is negative (see the blue line in figure ~\ref{fig:CCpAnalysis}) and the free surface profile of the locked wave and atmospheric pressure have the opposite signs. Thus, the positive atmospheric pressure induces a depression (negative elevation) in the locked wave free surface profile.
It is noted that for $F_r <1$, the free wave becomes the leading wave 

Applying the stationary phase approximation to $\eta_+$ in (\ref{solutions}), the far field solution can be expressed as, 
\begin
{equation}
    \eta_{+} \approx \frac{1}{2\rho g}\frac{1}{F_r-1}\left [-M_0\left (\frac{2}{\sqrt{gh}h^2 t} \right)^{1/3} A_i(Z)  +M_1\left (\frac{2}{\sqrt{gh}h^2 t} \right)^{2/3} A_i^\prime(Z) + \dots \right ]
\label{solution_trailing}
\end {equation}
where 
\begin{equation}
    Z=\left (\frac{2}{\sqrt{gh}h^2 t} \right)^{1/3}(x - \sqrt{gh}t); \;\;\; 
    M_0 = \int_{-\infty}^{\infty}P_a (x) dx \;\;\; \text{and}\;\;\; M_1 = \int_{-\infty}^{\infty}x P_a (x) dx
    \label{pressure area}
\end{equation} represent the area and the first moment under the atmospheric pressure curve, respectively.
In the solution, (\ref{solution_trailing}),  
$A_i(Z)$ is the Airy function and $A_i^\prime (Z)$ is its first derivative \citep[see figures~2.5 and 2.6 in][]{mei05pt1}.
Both functions are oscillatory for $Z < 0$ and decay exponentially for $Z > 0$.
However, $A_i(Z) > 0$ and $A_i^\prime (Z) < 0$ for $Z > 0$.
For the case where the first term dominates ($|M_0| > |M_1|$), the leading free waves decay as $t^{-1/3}$.
However, in the case where $M_0 = 0$ (e.g., the atmospheric pressure distribution has the shape of an isosceles $N$-wave), the free wave is represented by the second term in (\ref{solution_trailing}), which decays faster as $t^{-2/3}$. Finally, the sign of $\eta_+$ depends on the sign of  $M_0$ and whether it is under supercritical or subcritical condition.  

According to \cite{lynett22}, the atmospheric pressure for the Tonga event takes an $N$-wave shape that is not isosceles and  $M_0 > 0$. The order of magnitude of the ratio of the locked wave amplitude to that of the free wave at the far field can be estimated from (\ref{solution_leading}) and (\ref{solution_trailing}) as:
\begin{equation*}
\frac{O(\eta_{p})}{O(\eta_{+})} =O\left (-\frac{2^{5/3}}{(F_r+ 1)}\frac{P_a^c}{M_0}\left (\frac{1}{\sqrt{gh}h^2 t} \right)^{-1/3}
\right ),
\end{equation*}
where $P_a^c$ denotes the crest value of $P_a$ and   $O(A_i(Z)) \approx 1/2$ has been applied.
Denoting $S = \sqrt{gh} t$ as the distance that the front of the trailing wave has traveled at time $t$, the equation above can be simplified as 
\begin{equation}
\frac{O(\eta_{p})}{O(\eta_{+})} =O\left (-\frac{P_a^c h}{M_0}\left (2^5\frac{S}{h} \right)^{1/3} (F_r + 1)^{-1}
\right ).
\label{Ratio}
\end{equation}
The influence of the free waves diminishes as the atmospheric pressure wave propagates to infinity, i.e., $S \rightarrow \infty$.
For the free wave amplitude to be the same order of magnitude as the leading wave, $O(\eta_{p})/O(\eta_{+}) = O(1)$, the traveling distance of the free wave must be within the following relative distance, 
\begin {equation}
\frac{S}{h} < \frac{1}{2^5}\left (\frac{M_0}{P_a^c h}\right)^3(F_r +1 )^3. 
\end{equation}
These points will be further illustrated with DART data captured during the 2022 Tonga event in section~\ref{sec:applicationPacific}.

\subsection{Shallow water wave solutions}
\label{sec:LSWsol}
When the horizontal length scale of the atmospheric pressure wave is very long in comparison with the water depth, the generated water waves are non-dispersive long waves.  The simplified solutions can be readily deduced from (\ref{solutions}) by setting $C \rightarrow \sqrt{gh}$. Thus, the free surface shallow water wave solutions can be expressed as 
\begin {linenomath}
\begin {align}
\begin {split}
    \eta &=\eta_p +\eta_+ + \eta_-; \\
    \eta_p &= \frac{1}{\rho g}\frac{1}{F_r^2 -1} P_a(x-C_pt); \\
    \eta_+ &= -\frac{1}{\rho g}\frac{1}{2(F_r-1)}P_a(x-\sqrt{gh}t);\\
    \eta_- &= \frac{1}{\rho g}\frac{1}{2(F_r+1)}P_a(x+\sqrt{gh}t).
    \label{solutions_shallow}
\end{split}
\end{align}
\end{linenomath}

The solutions above, which can also be found in \cite{pelinovsky01}, satisfy the linear shallow water wave equations,
\begin{equation}
    \frac{\partial^2 \eta}{\partial t^2} - gh\frac{\partial^2 \eta}{\partial x^2} = \frac{gh}{\rho g} \frac{\partial^2 P_a}{\partial x^2}.
    \label{wave} \end{equation}
with the assumption that wave motions start from the quiescent condition.
Note that similar solutions for waves generated by a moving obstacle (e.g., landslide, ship) have also been obtained \citep{tinti01,lo21}.

Lacking the frequency dispersion, the resulting wave patterns are much simpler and are easier to interpret.
Moreover, most of the descriptions provided in section~\ref{sec:analysisDispersive} remain valid, as captured in figure~\ref{fig:CCpAnalysis} (note that the horizontal axis represents $F_r$ in this case).
The ratio between the free wave and the locked wave can now be calculated as
\begin{equation}
    O\left (\frac{\eta_p}{\eta_+}\right) = 
    \frac{-2}{F_r + 1}.
    \label{lockToFree}
\end{equation}
For the subcritical condition ($F_r < 1$), the locked wave is always larger than the free wave, up to a factor of 2 when $F_r \rightarrow 0$;
for the supercritical condition ($F_r > 1$) the free wave becomes larger than the locked wave.

\section{2DH axisymmetric shallow water wave problem}
\label{sec:LSWAxisymmetricsol}
In the Tonga event, the atmospheric pressure is nearly axisymmetric and decays in the radial direction \citep{amores22, lynett22}. In this section approximate solutions are sought after for axisymmetric shallow water waves, being forced by an atmospheric pressure field.  Thus, in terms of the free surface elevation, $\eta(r,t)$, the governing equation is well-known: 
\begin{equation*}
    \frac{\partial^2 \eta}{\partial t^2} - gh \frac{1}{r}\frac{\partial}{\partial r}\left( r \frac{\partial  \eta}{\partial r} \right ) = \frac{h}{\rho} \frac{1}{r}\frac{\partial}{\partial r}\left( r \frac{\partial  P_a}{\partial r} \right ),
    \label{wave_r}
\end{equation*}
which can be rewritten in the following form: 
\begin{equation}
    \frac{\partial^2 \sqrt{r} \eta}{\partial t^2} - gh\left(\frac{\partial^2 \sqrt{r}\eta}{\partial r^2} +  \frac{\sqrt{r}\eta}{4r^2}\right) = \frac{h}{\rho}\left(\frac{\partial^2 \sqrt{r}P_a}{\partial r^2} +  \frac{\sqrt{r}P_a}{4r^2}\right)
\end{equation}
Considering ``$l$'' as the characteristic length scale of the atmospheric pressure and the induced water wave. For large $r\gg l$, the second term, relative to the first term inside the brackets of the equation above, is  $ O (l/r)^2 << 1$ and can be neglected, resulting in an approximate governing equation in the far field as
\begin{equation}
    \frac{\partial^2 \sqrt{r} \eta}{\partial t^2} - gh\left(\frac{\partial^2 \sqrt{r}\eta}{\partial r^2}\right) = \frac{h}{\rho}\left(\frac{\partial^2 \sqrt{r}P_a}{\partial r^2}\right)
    \label{approxSWE}
\end{equation}
 Assuming that the atmospheric pressure takes the following form
\begin {equation}
P_a = \sqrt{\frac{r_0}{r}} P_0(r-C_pt),
\end{equation}
where $r_0$ is a constant, defining the radial location at which $P_a = P_0(r_0-C_pt)$, the analytical solution for (\ref{approxSWE}) can be obtained as 
\begin {equation}
\eta =\frac{1}{\rho g}\frac{1}{F_r^2-1}\sqrt{\frac{{r_0}}{{r}}}\left [ P_0(r-C_pt) - P_0(r-\sqrt{gh} t) \right ].
\label{eq:axisymmetricSolution1}
\end {equation} 
This result can also be  obtained by summing up the wave components of the 1DH solutions presented in (\ref{solutions_shallow}), and multiplying the resulting expression by the radial decay factor, $\sqrt{ r_0/r}$, since $\eta_{-}$ also propagates in the $ r-$direction (due to radial symmetry).  
\section{Applications to the 2022 Tonga event}
\label{sec:applicationPacific}

\begin{figure}
    \centering
    \includegraphics[width=0.39\linewidth]{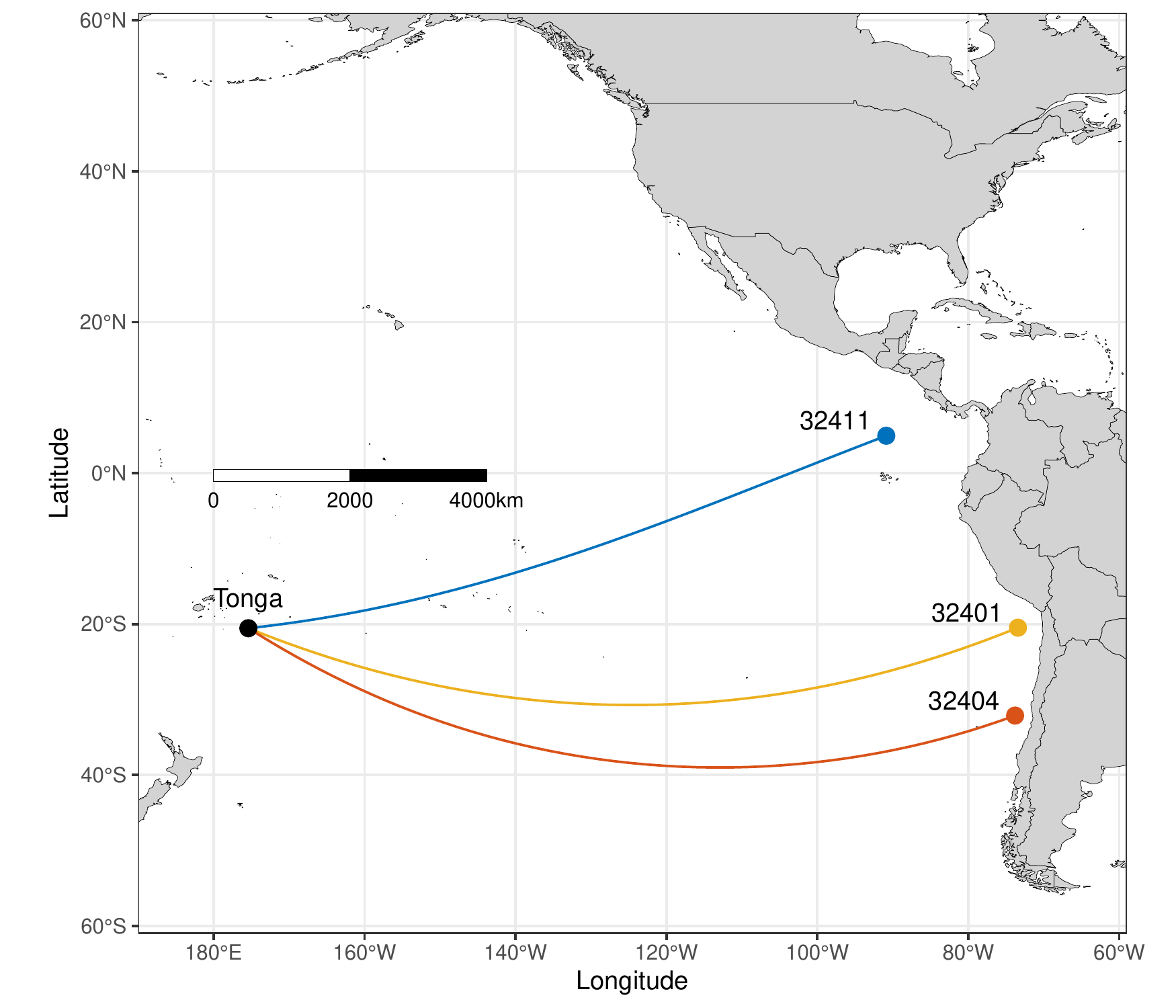}
    \includegraphics[width=0.60\linewidth,trim={3cm 0 3cm 0},clip]{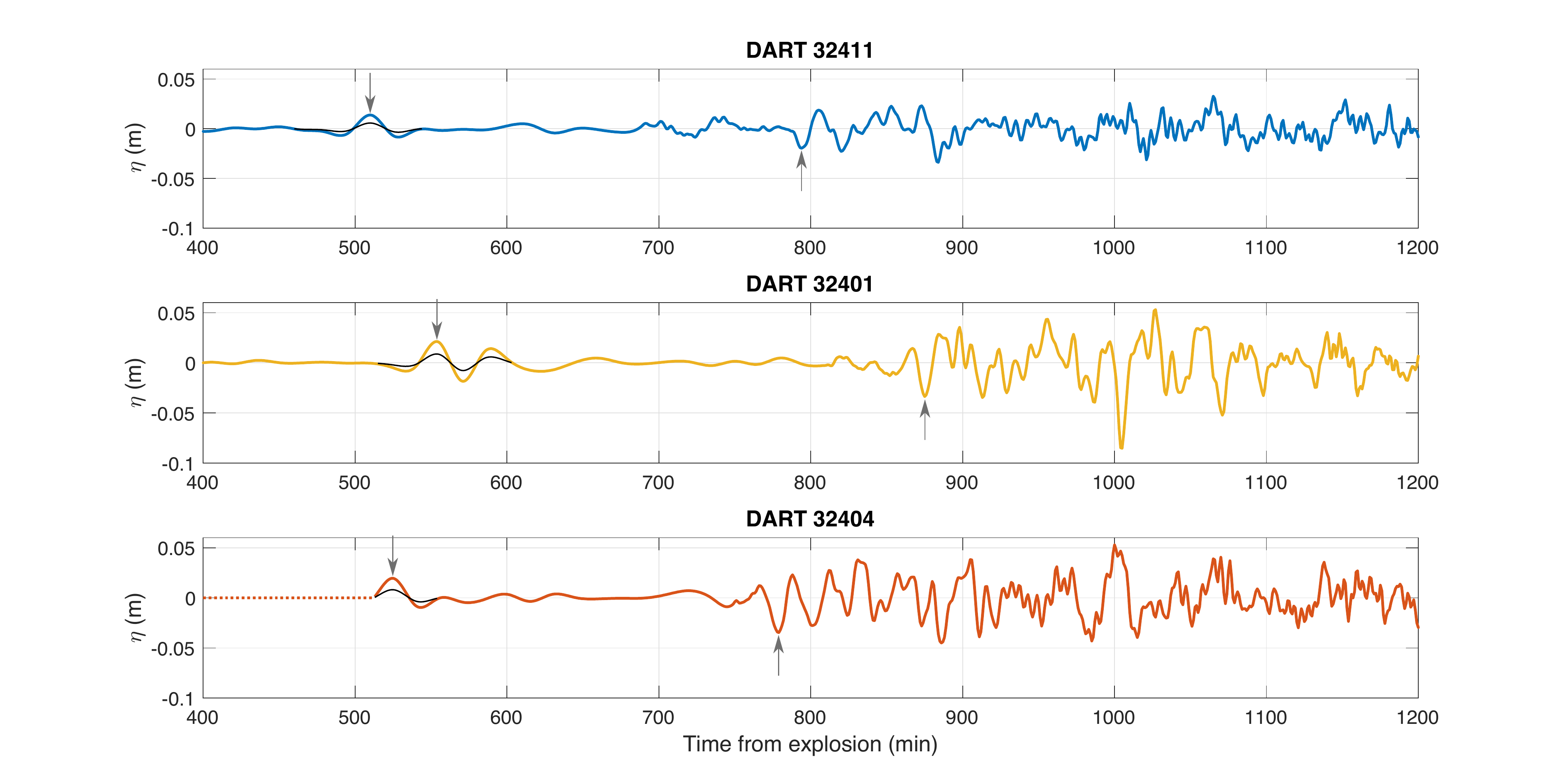}
    \caption{Geographical locations and time series of free surface elevation reported by DART stations. 
    The free surface elevations corrected by (\ref{eq:DARTcorrection}) are in black lines. Gray arrows mark the arrival times of the leading locked waves and trailing free waves.}
    \label{fig:DARTtimeSeries}
\end{figure}

The theoretical far-field solutions are used to check the three DART stations measurements (32411, 32404 and 32401) during the Tonga event, shown in the left panel in figure~\ref{fig:DARTtimeSeries}. The paths for the tsunamis reaching these stations are  practically uninterrupted from the source.
DART stations measure dynamic pressure at the bottom of the ocean.
Normally DART data is reported every 15 minutes, and when the system detects a tsunami, the reported data resolution is improved to every 15 seconds.
The reported data, $\zeta$, is calculated as follows  \citep{rabinovich15}:
\begin{equation}
\label{eq:etaDART}
\zeta = \eta + \frac{P_a}{\rho g},
\end{equation}
\noindent
capturing both the atmospheric pressure disturbances and the induced water waves for the leading (locked) wave. These data need to be corrected to identify the actual water wave surface profile, $\eta$.
For the Tonga event, the atmospheric pressure wave is long ($\sim 800$ km) and propagates within the supercritical regime ($F_r \approx 1.5$)\citep{amores22, lynett22}.
Thus, the free surface elevation of the leading locked wave has the same sign and shape as the atmospheric  pressure wave, and (\ref{solutions_shallow}) can be used in (\ref{eq:etaDART}) to find the following relationships:

\begin{equation}
\label{eq:DARTcorrection}
P_a = \rho g \left( \frac{F_r^2 - 1}{F_r^2} \right) \zeta, \;\;\; \text{and} \;\;\;
\eta =  \frac{1}{F_r^2}\zeta.
\end{equation}
This implies that the actual free surface elevation is smaller than the reported DART data, since $F_r^2 >1$. 
In addition, the first expression in (\ref{eq:DARTcorrection}) provides a formula for estimating the magnitude of the atmospheric pressure at the DART station, using the reported $\zeta$.
The time series shown in the right panels of figure~\ref{fig:DARTtimeSeries} contain both the reported DART data and the corrected data (black line) as per (\ref{eq:DARTcorrection}).

Practical information for the DART stations, such as the distance to Tonga, the average depth along the path, etc., are listed in table~\ref{table:pacificDART}.
The arrival times for the leading and trailing waves are marked with gray arrows in figure~\ref{fig:DARTtimeSeries} and listed  along with the separation times (in the last column) in table~\ref{table:pacificDART}.  Based on the DART data at these stations, the Froude numbers range from 1.48 to 1.58, with an average of 1.54, confirming that the Tonga event is under the supercritical condition.

During the event, the atmospheric pressure wave travels at the average velocity of $C_p \approx 1100$ km/hr in the Pacific Ocean \citep{lynett22}, and the wave celerity of long water waves can be estimated as $C \approx 713$ km/hr, corresponding to an average depth of $4$ km, representative of the Pacific Ocean. 
The theoretical time differences in the arrival times of the leading  locked wave and the trailing free wave are listed in the second to last column in table~\ref{table:pacificDART}.
The differences between the theoretical and observed time lapses are below 10\% for DART stations 32411 and 32401.
The difference for station 32404 is larger, as the trailing (free) waves arrive 37 minutes faster than expected. 

\begin{table}
\centering
\begin{tabular}{ c c c c c c c c c c c c c c }
 \textbf{Station} & $\bm{S}$ & $\bm{\overline{h}}$ & $\bm{t_{\eta_p}}$ & $\bm{t_{\eta_+}}$ & $\bm{\overline{C_p}}$ & $\bm{\overline{C}}$ & $\bm{\overline{F_r}}$ & $\bm{A_{\eta_p}}$ & $\bm{A'_{\eta_p}}$ & $\bm{P_{\eta_p}}$ & $\bm{A_{\eta_+}}$ & $\bm{\Delta t_{\mbox{\tiny theo}}}$ & $\bm{\Delta t_{\mbox{\tiny obs}}}$ \\ \hline
 32411 & 9,633 & 4,283 & 510 & 794 & 1133 & 728 & 1.56 & 1.38 & 0.57 & 0.81 & -1.96 & 285 & 284 \\
 32401 & 10,385 & 4,153 & 554 & 875 & 1125 & 712 & 1.58 & 2.13 & 0.85 & 1.28 & -3.36 & 307 & 321 \\
 32404 & 9,833 & 3,983 & 525 & 779 & 1124 & 757 & 1.48 & 1.96 & 0.89 & 1.07 & -3.46 & 291 & 254
\end{tabular}
\caption{Basic information from DART stations. $\bm{S}$:
great-circle distance from Tonga to the station in km; $\overline{\bm{h}}$:
average depth along the path in m; $\bm{t_{\eta_p}}$ and  $\bm{t_{\eta_+}}$:
arrival times of $\bm{\eta_p}$ and $\bm{\eta_+} $ in min; $\overline{\bm{C_p}}$ and  $\overline{\bm{C}}$:
average celerity of $\bm{\eta_p}$ and $\bm{\eta_+}$ in km/hr; $\overline{\bm{F_r}}$:
average Froude number; $\bm{A_{\eta_p}}$ and $\bm{A'_{\eta_p}}$:
wave amplitudes of $\bm{\eta_p}$ and their corrections, (\ref{eq:DARTcorrection}), in cm; $\bm{P_{\eta_p}}$:
estimated peak pressure, (\ref{eq:DARTcorrection}), in  hPa;
$\bm{A_{\eta_+}}$: wave amplitude of $\bm{\eta_+}$ in cm; $\bm{\Delta t_{\mbox{\tiny theo}}}$ and  $\bm{\Delta t_{\mbox{\tiny obs}}}$: time differences between the leading and trailing wave arrival times, from theoretical results and observations, in min.}
 
\label{table:pacificDART}
\end{table}

\begin{table}
\centering
\begin{tabular}{ c c c c c c }
 \textbf{Station} & \textbf{DART} & \textbf{CDART} & \textbf{(\ref{Ratio})} & \textbf{(\ref{lockToFree})} & \textbf{(\ref{eq:axisymmetricSolution1})} \\ \hline
 32411 & -0.70 & -0.29 & -1.03 & -0.78 & -1 \\
 32401 & -0.63 & -0.25 & -0.96 & -0.78 & -1 \\
 32404 & -0.57 & -0.26 & -0.97 & -0.81 & -1
\end{tabular}
\caption{Comparison of the $\eta_p/\eta+$ ratio for the uncorrected (DART) and corrected (CDART) data, and the analytical solutions.}
\label{table:DARTComparison}
\end{table}
The observed and the corrected  amplitudes of the leading and trailing waves, and the estimated peak atmospheric pressure according to (\ref{eq:DARTcorrection}) are also recorded in table~\ref{table:pacificDART}.
The peak pressures among these three DART stations range from 0.81 hPa to 1.28 hPa, with an average of 1.05 hPa, which is very close to 1.11 hPa, the value provided by the empirical model in \cite{lynett22}.

In table~\ref{table:DARTComparison}  the values of  $\eta_p/\eta+$ at each station are listed, including the reported and corrected DART data and various analytical solutions.
The ratio is always negative for all cases, indicating that the leading and trailing waves have opposite sign. All the analytical solutions show that the amplitude ratios are close to one, indicating that the shallow water wave theory is adequate in describing this event. 
As expected, applying the correction, (\ref{eq:DARTcorrection}), reduces the amplitude of the leading locked wave, thus, also decreases the value of the $\eta_p/\eta+$ ratio from an average of -0.63 to -0.27.
These corrected values are significantly smaller than any of the analytical values. Nevertheless, the measured data and analytical solutions are in agreement in the order of magnitude. 
The differences between the measured data and  theoretical solutions reflect the complexity of the problem, which includes effects of bathymetry, Earth's curvature and additional wave generation mechanisms related to the volcano explosion, which will travel together as part of the trailing wave package.

\section{Concluding remarks} \label{conclusion}
The analytical expressions developed herein cover dispersive and non-dispersive solutions and can be applied to model water waves generated by atmospheric pressure disturbances travelling at super- and sub-critical speeds.
They provide significant insights on the resulting water wave characteristics and can be used as benchmarks for numerical models.
It is shown that the wave patterns generated under the supercritical condition are fundamentally different from those generated by a pressure disturbance propagating in a subcritical condition.
Under the supercritical conditions the atmospheric pressure disturbances induce a leading ``locked'' wave with the same sign, i.e., a positive atmospheric pressure generates an elevated wave.
Under the subcritical conditions the locked wave is trailing and has an opposite sign.
Moreover, in the case in which the pressure wave is a long wave, the resulting water wave will have its same shape.
Atmospheric pressure disturbances also generate ``free'' waves, whose sign and shape are also determined by the Froude number.

In general, bottom-mounted pressure gauge measurements related to the locked waves need to be corrected to account for the additional pressure variations caused by the atmospheric pressure disturbances, which can be significant in the near field of the volcano explosion.
The correction method (\ref{eq:DARTcorrection}) is simple and useful in instances when atmospheric pressure measurements are not available. 

In conclusion, the analytical theories presented in this paper can explain the positive leading wave observed during Tonga's event, which is locked to the atmospheric pressure wave and, thus, arrives faster than expected based on the long wave celerity. Trailing waves are also produced by the atmospheric pressure wave.
Nevertheless, since these are free waves propagating at the long wave celerity, any other long waves produced during the explosion \citep[e.g., mechanical blast, collapse of the caldera, etc.][]{lynett22} would also be travelling as part of the same wave package.

\section*{\bf\noindent{Acknowledgements}}
P. L.-F. Liu would like to acknowledge the National University of Singapore research grant (NRF2018NRF-NSFC003ES-002). This research was also supported in part by the Yushan Program, Ministry of Education in Taiwan.







\bibliography{bibliographyPaper}
\bibliographystyle{jfm}

\end{document}